\documentclass[showpacs,preprintnumbers,amsmath,amssymb,prb,superscriptaddress,twocolumn]{revtex4}

\usepackage{epsfig}
\usepackage{dcolumn}% Align table columns on decimal point
\usepackage{bm}% bold math
\usepackage{latexsym}

\usepackage{color}
\newcommand{\imag}{\mathrm i}

\begin{document}

\title{Magnetic field induced nutation of the exciton-polariton polarization\\ in (Cd,Zn)Te crystals}

\author{T.~Godde}
\affiliation{Experimentelle Physik 2, Technische Universit\"at Dortmund, 44221 Dortmund, Germany}
\author{M.~M.~Glazov}
\affiliation{Ioffe Physical-Technical Institute, Russian Academy of Sciences, 194021 St. Petersburg, Russia}
\author{I.~A. Akimov}
\affiliation{Experimentelle Physik 2, Technische Universit\"at Dortmund, 44221 Dortmund, Germany}
\affiliation{Ioffe Physical-Technical Institute, Russian Academy of Sciences, 194021 St. Petersburg, Russia}
\author{D.~R. Yakovlev}
\affiliation{Experimentelle Physik 2, Technische Universit\"at Dortmund, 44221 Dortmund, Germany}
\affiliation{Ioffe Physical-Technical Institute, Russian Academy of Sciences, 194021 St. Petersburg, Russia}
\author{H.~Mariette}
\affiliation{CEA-CNRS group "Nanophysique et Semiconducteurs", Institut N\'eel, CNRS \\
                and Universit\'e Joseph Fourier, 25 Avenue des Martyrs, 38042 Grenoble, France}
\author{M. Bayer}
\affiliation{Experimentelle Physik 2, Technische Universit\"at Dortmund, 44221 Dortmund, Germany}
\affiliation{Ioffe Physical-Technical Institute, Russian Academy of Sciences, 194021 St. Petersburg, Russia}

\date{\today}

\begin{abstract}
We study the polarization dynamics of exciton-polaritons propagating in sub-mm thick (Cd,Zn)Te bulk crystals using polarimetric time-of-flight techniques. The application of a magnetic field in Faraday geometry leads to synchronous temporal oscillations of all Stokes parameters of an initially linearly or circularly polarized, spectrally broad optical pulse of 150~fs duration propagating through the crystal. Strong dispersion for photon energies close to the exciton resonance leads to stretching of the optical pulse to a duration of 200$-$300~ps and enhancement of magneto-optical effects such as the Faraday rotation and the non-reciprocal birefringence. The oscillation frequency of the exciton-polariton polarization increases with magnetic field $B$, reaching 10 GHz at $B\sim 5$~T. Surprisingly, the relative contributions of Faraday rotation and non-reciprocal birefringence undergo strong changes with photon energy, which is attributed to a non-trivial spectral dependence of Faraday rotation in the vicinity of the exciton resonance. This leads to polarization nutation of the transmitted optical pulse in the time domain. The results are well explained by a model that accounts for Faraday rotation and magneto-spatial dispersion in zinc-blende crystals. We evaluate the exciton $g$-factor $|g_{\rm exc}|=0.2$  and the magneto-spatial constant $V= 5 \times 10^{-12}$~eVcm$\textup{T}^{-1}$.
\end{abstract}

\pacs{ 71.35.Ji/ 71.36.+c / 78.47.D- /78.20.Ls}

\keywords{exciton polariton dynamics, Magneto-spatial dispersion, Quantum beats}

\maketitle

\section{Introduction}
The propagation of light in an anisotropic medium can be characterized by the transformation of the polarization state of the transmitted electromagnetic wave compared to its initial polarization. This phenomenon has been widely used for intensity and phase modulation of light. External forces such as deformation or electric and magnetic fields can be used to influence the optical anisotropy of the medium.\cite{Zvezdin-book}
Fast switching and full control of the polarization state of light are essential for applications in rapidly developing fields of photonics such as quantum cryptography.\cite{Crypt02}
The evolution of the polarization state in an actively controlled anisotropic medium is also interesting from fundamental point of view because it is the equivalent to the evolution of an electron spin in a magnetic field.\cite{Zapasski99}
Several spectacular effects such as nutation or echo of the polarization have been demonstrated.\cite{Slichter-book}  These phenomena imply the coherence of the light. Therefore, first it is essential that this coherence is maintained during propagation. Second, in order to realize full control of the polarization state it is necessary to have tools for changing the superposition of the two eigenmodes of the electromagnetic waves. This means that, for example, if we use linear and circular birefringence to control the polarization state then the relative strength of their contributions should be variable. While the first issue regarding coherence is easy to fulfil in the transparency region of a material, realization of the second requirement needs more efforts.

This work is devoted to investigation of the magnetic field induced transformation of the polarization of light, that propagates through a (Cd,Zn)Te bulk semiconductor with photon energy close to the exciton resonance. The choice of material is specific for the following reasons.
First, because it has zinc blende structure, it belongs to the symmetry class $T_{\rm d}$ and lacks a center of inversion.
Hence, in addition to the classical magneto-optical Faraday and Voigt effects, it exhibits magneto-spatial dispersion, a combined effect involving the light wavevector $\bm{k}$ and the magnetic field $\bm{B}$.\cite{Excitons,Burstein71}
Second, we demonstrated recently that the propagation of light in a (Cd,Zn)Te crystal with photon energy below the exciton resonance can be well described in terms of exciton-polariton propagation.\cite{Godde10}
The ballistic propagation of exciton-polaritons from the lower branch with low absorption is slowed down up to 150 times depending on the photon energy.
Therefore, long-distance ($\sim 1$~mm) coherent propagation and sub-ns time delays of optical pulses were obtained for the light propagation. The latter is quite important because generally the strength of magneto-optical effects is proportional to the time delay introduced by the medium. This makes the arrival time of the pulse a representative scale for these effects.

The effect of magneto-spatial dispersion deserves special attention. In the transparency region it manifests as non-reciprocal birefringence, studied in several semiconductors such as wurtzite CdSe,\cite{Ivchenko84-CdSe} and cubic GaAs, CdTe and ZnTe.\cite{Gogolin84-GaAs,Krichevtsov99-CdTe}
Nonreciprocal birefringence was also demonstrated in the diluted magnetic semiconductor (Cd,Mn)Te.\cite{Krichevtsov98-CdMnTe}
Since it involves both light wavevector and external magnetic field, the magneto-spatial dispersion is, as a rule, much weaker than the Faraday rotation, and hence, to the best of our knowledge, all previous studies of non-reciprocal birefringence were done in Voigt geometry, where the Faraday effect is negligible, while the Voigt effect is quadratic in magnetic field and must be taken into account.

In this work we apply a polarimetric time-of-flight technique, which allows one to resolve the polarization state of the transmitted optical pulse in real time. We demonstrate that when approaching the exciton resonance in (Cd,Zn)Te crystals the Faraday effect disappears at a certain photon energy, making it possible to observe the non-reciprocal birefringence.
Such behavior occurs because the Faraday effect is governed by the superposition of two contributions with different signs.
These contributions result from the exciton resonance at the fundamental band gap edge in the center of the Brillouin zone ($\Gamma$-point) and from energetically higher lying resonances with much larger oscillator strengths, which correspond to interband optical transitions at the $X$ and $L$ points of the Brillouin zone with high densities of electronic states.\cite{Cade1985, Furdyna04} The non-reciprocal linear birefringence, on the other hand, is determined mainly by the single exciton resonance and its strength increases when the photon energy approaches the exciton resonance.
Such a combination of circular (Faraday effect) and linear (magneto-spatial dispersion effect) birefringence leads to the extraordinary case in which the optical anisotropy depends strongly on photon energy. We show that the complex interplay between linear and circular birefringence and dichroism results in a \emph{polarization nutation} effect in (Cd,Zn)Te crystals in the vicinity of the exciton resonance.

The manuscript is organized as follows. Section \ref{sec:theo} describes the theoretical basis of the exciton-polariton propagation and the accompanying polarization transformation due to magnetic-field-induced circular- and linear birefringence. The experimental details are given in Sec.~\ref{sec:Exp}. Section~\ref{sec:cw} describes the magneto-optical properties of (Cd,Zn)Te crystals evaluated from time-integrated reflection, transmission and photoluminescence measurements. Here we also evaluate the role of the inhomogeneous broadening of the exciton resonance.  Section~\ref{sec:TOF} describes the main time-resolved polarimetric measurements.
The spectral dependence of the relative strength of linear- and circular birefringence is presented in Sec.~\ref{sec:circ-lin}. Here we also present the complete picture of exciton-polariton propagation and polarization nutation accounting for Faraday rotation, non-reciprocal birefringence and circular dichroism in a wide range of magnetic fields up to 7~T.

\section{\label{sec:theo} Theoretical background}
\subsection{\label{subsec:pheno} Phenomenological model}

The optical properties of a crystal can be conveniently described by the dielectric susceptibility tensor $\varepsilon_{ij}$ whose Cartesian components ($i,j=x,y,z$) are functions of the light frequency $\omega$, the light wavevector $\bm k$ and the external magnetic field $\bm B$. They satisfy the restrictions imposed by the time reversal symmetry (Onsager principle)
\begin{equation}
\label{eps:gen}
\varepsilon_{ij}(\omega, \bm k, \bm B) = \varepsilon_{ji}(\omega, - \bm k, - \bm B),
\end{equation}
as well as the restrictions imposed by the point symmetry of the system, which in a (Cd,Zn)Te cubic crystal, is described by the $T_d$ point group. Far from the exciton resonance the dielectric susceptibility, Eq.~\eqref{eps:gen}, can be decomposed in powers of $\bm k$ and $\bm B$. By retaining only those linear in $\bm B$ and $k_iB_j$ (note, that $\bm k$-linear terms in $\varepsilon_{ij}$ are forbidden for the $T_d$ point group) we obtain\cite{Gogolin84-GaAs}
\begin{multline}
\label{epsilon1}
\varepsilon_{ij} = \varkappa \delta_{ij} +   \imag \gamma_1 \delta_{ijl} B_l \\
- \frac{\gamma_2}{2}\delta_{ij} (k_{i+1}B_{i+1} - k_{i+2}B_{i+2}) +\gamma_3 (k_i B_j - k_jB_i)\textup{.}
\end{multline}
Here a cubic axes frame is used ($\bm x\parallel [100]$, $\bm y\parallel [010]$ and $\bm z\parallel [001]$), the cyclic rule is applied ($i+3=i$), $\delta_{ij}$ is the Kronecker $\delta$-symbol, $\delta_{ijl}$ is the Levy-Civita symbol, $\varkappa$ is the dielectric susceptibility for $\bm k= \bm 0$, $\bm B= \bm 0$, $\gamma_1$ is a constant responsible for the Faraday effect and $\gamma_2$, $\gamma_3$ are constants responsible for the magneto-spatial dispersion.

Our experiments are carried out in the Faraday geometry, where $\bm B \parallel \bm k \parallel \bm z$. Then the relevant contributions to the dielectric tensor are given by
\begin{subequations}
\label{epsilon}
\begin{equation}
\label{farad}
\varepsilon_{xy} = - \varepsilon_{yx} = \imag \gamma_1 B_z,
\end{equation}
\begin{equation}
\label{disp}
\varepsilon_{xx}  = \varkappa + \frac{\gamma_2}{2} k_z B_z,\quad \varepsilon_{yy}  = \varkappa - \frac{\gamma_2}{2} k_z B_z.
\end{equation}
\end{subequations}
In this case, the effect of $\gamma_3$ vanishes, however, both circular birefringence/dichroism and linear birefringence/dichroism are present. As a result, the eigenmodes of the electromagnetic field in the crystal are, in general, neither linearly nor circularly polarized, but they are elliptically polarized. To illustrate this we neglect absorption (i.e. assume that $\varkappa$, $\gamma_1$ and $\gamma_2$ are real) and present the effective refractive indices of the eigenwaves as
\begin{equation}
\label{neff}
n_{\pm}^2=\varkappa \pm B_z \sqrt{\gamma_1^2 + q^2}.
\end{equation}
Here $q= \gamma_2 k_z/(2\gamma_1)$ characterizes the relative strengths of linear and circular birefringence. The corresponding eigenmodes have the following polarization vectors
\begin{equation}
\label{e}
\bm e_\pm = \frac{1}{2\sqrt{1+q^2}(\sqrt{1+q^2} \pm q)} \left( \imag q \pm  \imag \sqrt{1+q^2}, 1 \right).
\end{equation}
As a result, the initially linearly polarized radiation after transmission through such a medium becomes elliptically polarized with the degree of circular polarization and the orientation of the polarization ellipse determined by the optical path. The phase difference between the orthogonally polarized eigenwaves transmitted through the crystal is given by
\begin{equation}
\label{phase:diff:gen}
\phi = (n_+ - n_-)\frac{\omega L}{c}\approx  B_z \sqrt{\gamma_1^2+q^2}\frac{\omega L}{\sqrt{\varkappa}c},
\end{equation}
where the approximate equality holds for $|n_+ - n_-|\ll n_+, n_-$.

\subsection{\label{subsec:model}Magnetic-field-induced optical anisotropy in the exciton region}

In the vicinity of the exciton the components of the dielectric susceptibility tensor contain resonant features, so that the power series decomposition in Eq.~\eqref{epsilon} is no longer valid, in general.\cite{Excitons} Therefore, it is instructive to relate the components $\varepsilon_{ij}(\omega,\bm k, \bm B)$ with the parameters of the exciton in bulk. To that end, we present Schr\"{o}dinger equations for the exciton wavefunction in the form\cite{ivch:excitons}
\begin{equation}
\label{exc:gen}
\left[ \mathcal H_{\nu\nu'}(\bm k) - \hbar\omega \delta_{\nu\nu'} -
  \mathrm i\hbar \Gamma \delta_{\nu\nu'}\right] C_{\nu'}(\bm k) =\bm
d_{\nu} \cdot \bm E({\bm k,\omega}).
\end{equation}
Here $\bm E$ is the electric field of the electromagnetic wave propagating inside the crystal, $\nu = \sigma^{+}$ or $\sigma^{-}$ denotes exciton states which are excited by correspondingly polarized photons, $C_{\nu}$ are the components of the exciton wavefunction, $\mathcal H(\bm k)$ is the exciton
Hamiltonian, and $\bm d_\nu$ is the exciton dipole moment matrix
element. We have also included a phenomenological damping,
$\Gamma$, into
Eq.~\eqref{exc:gen}. Neglecting the exciton dispersion (limit of infinite effective mass) we obtain
the following $2\times 2$ effective Hamiltonian (for $\bm k \parallel \bm B \parallel [001]$:
\begin{equation}
\label{Heff}
\mathcal H =
\begin{pmatrix}
E_{\rm exc} + \frac{\Delta}{2} & V k_z B_z \\
V k_z B_z & E_{\rm exc} - \frac{\Delta}{2}
\end{pmatrix},
\end{equation}
describing the doublet of optically active states.
Here $E_{\rm exc}$ is the exciton resonance energy, $\Delta = g_{\rm exc} \mu_B B_z$ is the exciton Zeeman
splitting with $g_{\rm exc}$ being its $g$-factor and $\mu_B$ the Bohr
magneton, and the constant $V$ is responsible for the magneto-spatial dispersion. Calculating the exciton resonance contribution to the dielectric polarization, $\bm P_{\rm exc} = \sum_{\nu} \bm d_{\nu} C_{\nu}$, we finally obtain for the components of the dielectric susceptibility tensor
\begin{subequations}
\label{epsilon:exc}
\begin{equation}
\label{e:xx}
\varepsilon_{xx} = \varepsilon_b + \frac{4\pi |d|^2}{\mathcal D} (E_{\rm exc} -
\hbar\omega - \mathrm i \hbar \Gamma - V k_z B_z),
\end{equation}
\begin{equation}
\label{e:yy}
\varepsilon_{yy} = \varepsilon_b + \frac{4\pi |d|^2}{\mathcal D} (E_{\rm exc} -
\hbar\omega - \mathrm i \hbar \Gamma + V k_z B_z),
\end{equation}
and
\begin{equation}
\label{e:xy}
\varepsilon_{xy} = -\varepsilon_{yx} = -\frac{4\pi |d|^2}{\mathcal D}
\frac{\Delta}{2\mathrm i} = \mathrm i \Delta \frac{2\pi |d|^2}{\mathcal D}.
\end{equation}
\end{subequations}
Here $\varepsilon_b$ is the background dielectric constant,
\begin{equation}
\label{Den}
\mathcal D = (E_{\rm exc} - \frac{\Delta}{2} - \hbar\omega - \mathrm i
\hbar \Gamma) (E_{\rm exc} + \frac{\Delta}{2} - \hbar\omega - \mathrm i
\hbar \Gamma) - (V k_z B_z)^2 \textup{.}
\end{equation}
We note, that $4\pi |d|^2$ can be rewritten as $\varepsilon_b \hbar
\omega_{LT}$, where $\hbar\omega_{LT}$ is the longitudinal-transverse
splitting. In  bulk CdTe $\varepsilon_b=11.2$ and $\hbar\omega_{LT}=0.65$~meV.\cite{Sooryakumar83}

The dielectric tensor in Eqs.~(\ref{epsilon:exc}) describes the main magneto-optical effects resulting from the exciton resonance, in particular, magnetic-field induced circular birefringence and dichroism, linear birefringence as well as dichroism caused by the magneto-spatial dispersion. The parameters of the decomposition~\eqref{epsilon} can be expressed as
\begin{subequations}
\label{eq:gammas}
\begin{equation}
\label{kappa}
\varkappa = \varepsilon_b + \frac{4\pi |d|^2}{\mathcal D} (E_{\rm exc} - \hbar \omega - \mathrm i \hbar \Gamma)
\end{equation}
\begin{equation}
\label{subeq:gamma}
\gamma_1=\frac{2\pi|d|^2}{\mathcal D}\frac{\Delta}{B_z}, \quad
\gamma_2=-\frac{8\pi|d|^2}{\mathcal D}\tilde{V}.
\end{equation}
\end{subequations}
We stress that the components of the dielectric tensor depend, in general, non-linearly on the magnetic field and the wavevector. However, away from the exciton resonance $|E_{\rm exc} - \hbar\omega| \gg \hbar \Gamma$, or if $|\Delta| \ll \hbar \Gamma$, $|V k_z B_z|\ll \hbar \Gamma$,  the dependence of the denominators $\mathcal D$ in Eqs.~\eqref{eq:gammas} on $\bm k$ and $\bm B$ can be neglected. In this case, the parameters $\gamma_1$ and $\gamma_2$ do not depend on $k_z$ and $B_z$ and decomposition Eq.~\eqref{epsilon1} holds.

For sake of illustration we consider the transparency region, where absorption effects can be disregarded. Then the phase difference between the two eigenwaves of the exciton-polariton accumulated after propagating a distance $L$ is defined as
\begin{equation}
\label{eq:phi}
\phi = \frac{\omega}{c} L (n_+-n_-)\approx \frac{\omega}{c}L \frac{\varepsilon_b \hbar\omega_{\rm LT}}{2 n_0 \mathcal D} \hbar\Omega_{\rm eff},
\end{equation}
where $n_0=\sqrt{\varkappa}$ is the frequency dependent refractive index at $B=0$ (the latter includes the exciton contribution), and
\[
\hbar\Omega_{\rm eff}=\sqrt{\Delta^2+(2k_zB_z V)^2},
\]
is the effective splitting of the exciton resonance, which is a linear function of $B_z$. Equation~\eqref{eq:phi} was obtained assuming, that the magnetic field induced contributions to the dielectric tensor are small compared to $\varkappa$, i.e. $\varkappa \gg B_z \sqrt{\gamma_1^2 + ({\gamma_2 k_z}{/2})^2}$.
Moreover, for our experimental conditions, the Zeeman splitting is much smaller than the photon energy detuning with respect to the exciton resonance ($E_{\rm exc}-\hbar\omega \gg \Delta$), which is also true for the contributions $|k_zB_z V| \ll E_{\rm exc}-\hbar\omega$.
Hence, the magnetic field terms in  Eq.~\eqref{Den} are neglected and $\mathcal D = (E_{\rm exc}-\hbar\omega- \rm i\hbar \Gamma)^2$.
Then in the transparency region the phase difference can be written as [cf. Eq.~\eqref{phase:diff:gen}]
\begin{equation}
\label{eq:phi-group}
\phi = \omega \frac{d n_0}{d\omega} \frac{L}{c} \Omega_{\rm eff}=(n_g-n_0)\frac{L}{c}\Omega_{\rm eff},
\end{equation}
where
\begin{equation}
\label{eq:n_group}
n_g= \frac d {d \omega}Re[ \omega \sqrt{\varkappa(\omega)}],
\end{equation}
is the group refractive index, which defines the group velocity of exciton-polariton propagation. The above expression shows that the phase difference is determined mainly by the delay of the light pulse due to exciton-polariton propagation $\tau=(n_g-n_0)L/c$. In our previous work, Ref.~\onlinecite{Godde10} we demonstrated that exciton-polariton delays can be large due to the significant increase of the group index (up to 100) when approaching the exciton resonance.

To describe qualitatively the transformation of the polarization one has to take into account the absorption effects, which become especially important when approaching the exciton resonance.  To that end, it is necessary also to take into account the inhomogeneous broadening of the exciton resonance.
In this case the dielectric function can be found by convoluting Eqs. \eqref{epsilon:exc} with a Gaussian distribution:
\begin{equation}
\label{eq:kappa}
\varkappa(\omega)=\varepsilon_b + \int_{-\infty}^{+\infty} \frac{\varepsilon_b \omega_{LT}}{\omega^\prime - \omega - \mathrm i \Gamma} \frac{\exp{\left[ -\frac{(\omega^\prime-\omega_{0})^2}{2 \Gamma_{\rm inh}^2}\right]}} {\sqrt{2 \pi} \Gamma_{\rm inh}} d\omega^\prime \textup{, }
\end{equation}
where $\Gamma_{\rm inh}$ is the inhomogeneous broadening and $\hbar\omega_0$ corresponds to the central energy of the exciton resonance. The experimental results on the spectral dependence of transmission in Section~\ref{sec:cw} indicate that the homogenous linewidth of the exciton resonance is much smaller than the inhomogenious width ($\Gamma_{\rm inh}\gg\Gamma$).

\subsection{\label{subsec:micro}Possible microscopic origins of linear birefringence}

The presence of magnetic-field-induced spatial dispersion is related to the lack of an inversion center in the zinc-blende lattice of the (Cd,Zn)Te crystal. In Ref.~\onlinecite{Gogolin84-GaAs} it was assumed that it is caused by the  spin-dependent terms linear in the hole wavevector $\bm K$, that are allowed in structures of $T_d$ point symmetry,\cite{pikusmarushaktitkov}
\begin{equation}
\label{k-lin}
\mathcal H_1 = \frac{4}{\sqrt{3}} k_0 (K_x \mathcal V_x + K_y \mathcal V_y + K_z \mathcal V_z),
\end{equation}
where $k_0$ is a constant, $\mathcal V_x = \{J_x(J_y^2-J_z^2)\}_{\rm sym}$, etc., and $J_x$, $J_y$, $J_z$ are the angular momentum $3/2$ matrices relevant for the $\Gamma_8$ valence band. The corresponding contribution to $V$ can be estimated as\cite{Gogolin84-GaAs}
\begin{equation}
\label{tildeV:Klin}
V \sim \frac{\mu_B}{\mathcal R} k_0 \frac{m_0}{M},
\end{equation}
where $\mathcal R$ is the exciton Rydberg, $M$ is the translational mass of the exciton, $m_0$ is the free electron mass and numerical coefficients of the order of unity are omitted. For GaAs crystals the comparison of $V$ in Eq.~\eqref{tildeV:Klin} with experimental data of magnetic-field-induced spatial dispersion gives $k_0 \approx 30$~meV\AA~which is about an order of magnitude larger than the microscopic estimate of $k_0\approx-1.7$~meV\AA~given in Ref.~\onlinecite{pikusmarushaktitkov}, see also Ref.~\onlinecite{note1}.

Besides $K$-linear terms the hole Hamiltonian contains $K^3$ contributions\cite{pikusmarushaktitkov} and, what is more important, contributions linear in $\bm K$ and $\bm B$. A symmetry analysis shows that the following terms are allowed for the $\Gamma_8$ band
\begin{equation}
\label{G8:phen}
\mathcal H_2 = \alpha K_z B_z (J_x^2-J_y^2) + {\rm c.p.},
\end{equation}
where $\alpha$ is a constant and c.p. denotes the cyclic permutations. These contributions result in magnetic-field-induced mixing of the light- and heavy-holes for $\bm K\ne 0$ and, eventually, mix the optically active exciton states. The latter are formed from the electron ($\Gamma_6$ band, spin components $s_z =\pm 1/2$) and combinations of light- and heavy-holes ($\Gamma_8$ band, momentum components $\pm 1/2$ or $3/2$, respectively):
\begin{subequations}
\label{psi:x}
\begin{equation}
|+1\rangle = \frac{\sqrt{3}}{2} |\Gamma_8,3/2; \Gamma_6, -1/2\rangle -
\frac{1}{2} |\Gamma_8,1/2;\Gamma_6,1/2 \rangle,
\end{equation}
\begin{equation}
|-1\rangle =-\frac{\sqrt{3}}{2} |\Gamma_8,-3/2; \Gamma_6, 1/2\rangle +
\frac{1}{2} |\Gamma_8,-1/2;\Gamma_6,-1/2 \rangle.
\end{equation}
\end{subequations}

In order to estimate the constant $\alpha$ and, correspondingly, the magneto-spatial dispersion we resort to the extended (8-band) Kane model, where the lack of an inversion center is taken into account as $K^2$ off-diagonal terms in the $\bm k\cdot\bm p$ Hamiltonian.\cite{note2} The mixing term has the form
\begin{widetext}
\begin{equation}
\label{Hmm1}
\langle -1/2, \Gamma_8 | \mathcal H_{\bm k \bm p } | 3/2, \Gamma_8\rangle = \sum_c \frac{\mathcal H_{\Gamma_8 -1/2,c} \mathcal H_{c,\Gamma_8 3/2}}{E_v - E_c} = -\frac{\hbar^2}{m_0^2 E_g} \sum_c (\bm \kappa^*\bm p_{\Gamma_8 -1/2,c})(\bm \kappa\bm p_{c,\Gamma_8 3/2}),
\end{equation}
\end{widetext}
where $c$ enumerates the conduction band states, $\bm p_{cm}$ ($m=\Gamma_8 3/2$, \ldots) are the matrix elements of the momentum operator and
\[
\kappa_i =  K_i - \mathrm i \beta  K_{i+1} K_{i+2}.
\]
$\beta$ is the constant which takes into account the $\bm k\bm p$  admixture of the remote conduction band. It is related to the parameter $m_{\rm cv}$ introduced in Refs.~\onlinecite{pikusmarushaktitkov,OO,ivchenkopikus} by
\begin{equation}
\label{beta}
\beta = \frac{m_0}{m_{\rm cv}} \frac{\hbar}{ p_{\rm cv}},
\end{equation}
where $p_{\rm cv}$ is the interband momentum matrix element. Taking into account the presence of a magnetic field
 and making use of the fact that
\[
\hat K_x \hat K_y - \hat K_y \hat K_x = \frac{\mathrm i e}{\hbar c} B_z,
\]
we obtain the linear-in-$k_z B_z$ contribution in the form of Eq.~\eqref{G8:phen} with
\begin{equation}
\alpha = \frac{e\hbar |p_{\rm cv}|^2}{3 c m_0^2 E_g}\beta.
\end{equation}
Correspondingly,
\begin{equation}
\label{tildeV:KB}
V = \frac{3}{4} \alpha = \frac{e\hbar |p_{\rm cv}|^2}{4 c m_0^2 E_g}\beta.
\end{equation}
\section{\label{sec:Exp} Experimental details}
The investigated Cd$_{0.88}$Zn$_{0.12}$Te crystal was grown by the Bridgman technique at high temperature (1200$^\circ$C).
The crystal was cut along the (100) plane and divided into samples that were chemically-mechanically polished to different thicknesses in the range from 208~$\mu$m to 745~$\mu$m.
Further on, we focused on the 655~$\mu$m thick sample.
The samples had a very good sufarce quality (as evidenced by a rocking curve width of less than 20 arcsec, etched pits density of $10^4$~cm$^{-2}$ ).
The samples were mounted in a He-bath magneto-optical split-coil cryostat and maintained at a temperature of 1.8~K.
Magnetic fields up to 7~T in Faraday geometry with $\bm{B}  \| [001]$ were applied.

For time-of-flight measurements we used a mode-locked Ti:Sa laser as a source of Fourier transform limited optical pulses with duration of about 150~fs. The laser beam was focused on the sample at normal incidence along the $z$-axis ($\bm{k}\|[001]$).
The transmitted or reflected beam was dispersed by a single 0.5~m spectrometer with 6~nm/mm linear dispersion and detected by a streak camera, which was synchronized to the Ti:Sa laser operating at 75.75~MHz pulse repetition rate.
The overall temporal and spectral resolution of the experimental setup for time-resolved measurements was about 20 ps and 0.5 nm, respectively.
The time of arrival of the exciton-polariton $\tau$ is measured with respect to the time of arrival of the optical pulse propagating through the same apparatus without the sample.

The polarization of the incident and transmitted (reflected) beams was selected by means of Glan-Thompson prisms in conjunction with half- or quarter-wave plates. This allows us to determine all three Stokes parameters for the detected signal. These parameters are defined as
\begin{equation}
\label{eq:Stokes}
S_1=\frac{I_x - I_y}{I_x + I_y},\quad S_2=\frac{I_{x^\prime} - I_{y^\prime}}{I_{x^\prime} + I_{y^\prime}},\quad S_3=\frac{I_{\sigma^+} - I_{\sigma^-}}{I_{\sigma^+} + I_{\sigma^-}},
\end{equation}
where $I_x$, $I_y$, $I_{x^\prime}$, and $I_{y^\prime}$ are the intensities of the signal detected for linear polarization along $\bm x\parallel [100]$, $\bm y\parallel [010]$, $\bm x^\prime \| [110]$, and $\bm y^\prime \| [\bar{1}10]$, respectively. The intensities $I_{\sigma^+}$ and $I_{\sigma^-}$ correspond to $\sigma^+$ and $\sigma^-$ polarized detection, respectively. The orientation of the sample with respect to the laboratory coordinate system was confirmed by Laue X-ray diffraction.

For time-integrated measurements we used a single or triple spectrometer equipped with a liquid nitrogen-cooled charge coupled device camera or a single channel photomultiplier, respectively. Photoluminescence (PL) spectra were measured in back-scattering geometry under laser excitation with photon energy 2.33~eV and $\approx$~10~W/cm$^2$ power density. Steady-state reflectivity and transmission measurements were performed using a halogen lamp.

\section{\label{sec:cw} Diamagnetic shift and Zeeman splitting of exciton states}

\begin{figure}
\includegraphics[width=7.5 cm]{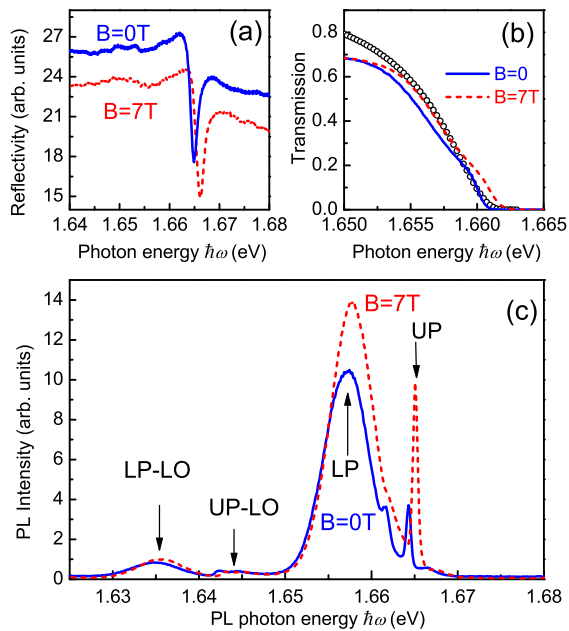}
\caption{(a) Reflection, (b) transmission, and (c) PL spectra measured at $T=1.8$~K of the (Cd,Zn)Te crystal taken at $B=0$ (solid blue lines) and 7~T (dotted red lines). PL spectrum is recorded for non-resonant excitation with photon energy $\hbar \omega_{ex}=2.33$~eV.
In addition a calculated transmission spectra for $B=0$ is shown in (b) assuming an inhomogeneously broadened system with $\hbar\Gamma=4~\mu$eV, $\hbar\Gamma_{\rm inh}=1$~meV (black circles).
The absorption coefficient is obtained using Eq.~\eqref{eq:kappa} with $\hbar\omega_{LT}=0.65$~meV and $\varepsilon_b=11.2$ from Ref. \onlinecite{Sooryakumar83} and $\hbar\omega_0=1.6644$~eV.
 \label{fig:PL}}
\end{figure}

Time-integrated reflectivity and PL spectra allow us to determine the diamagnetic shift and the Zeeman splitting of the exciton states.\cite{ivchenkopikus,Cho75,Dreybrodt80} Figure \ref{fig:PL} shows reflection (a), transmission (b) and PL spectra (c) taken at $B=0$ and 7~T. The reflectivity spectrum contains a dip, which is attributed to the exciton resonance.
At zero magnetic field the energy of this resonance is $\hbar\omega_0=1.6644$~eV. The spectral width of about 1~meV is due to inhomogeneous broadening.
In our case the origin of inhomogeneous broadening in (Cd,Zn)Te crystals is attributed mainly to fluctuations of composition in ternary alloys.\cite{Permogorov-Reznitsky}

The transmission spectra taken for a $655~ \mu$m thick crystal are shown in Fig.~\ref{fig:PL}(b).
For $B=0$~T the transmission intensity drops from $70$~\% at $1.65$~eV to values below $1$~\% above $1.66$~eV.
This allows transmission measurements in thick sub-mm crystals without significant absorption for photon energies up to $5$~meV below the exciton resonance (detuning energy $\delta = \hbar\omega_0 -\hbar \omega  \geq  5$~meV).
This is only possible due to a low homogeneous linewidth of $\hbar\Gamma=4~\mu$eV, which was determined by a fit of the transmission edge. The absorption coefficients used in the fit were calculated using Eq.~\eqref{eq:kappa} with $\hbar \Gamma_\textup{inh}=1$~meV, $\hbar\omega_0=1.664$~eV and $\hbar \omega_{LT} = 0.65$~eV.
Such narrow homogenous linewidth is not surprising for a exciton resonances, which are inherent to bulk material. Similar values $\hbar\Gamma\sim 10~\mu$eV have been recently reported in ZnO and GaN bulk crystals.\cite{Shubina-ZnO,Shubina-GaN,Shubina08}

The PL spectrum comprises  several peaks, which can be attributed to radiative emission of exciton-polaritons from the upper (UP peak at 1.6643~eV) and lower (broad LP peak at 1.657~eV) polariton branches and phonon replicas, related with radiation after emission of an LO phonon (UP-LO and LP-LO peaks), respectively. The detailed assignment of these features and their kinetics are presented in Ref.~\onlinecite{Godde10}.

\begin{figure}
\includegraphics[width=7 cm]{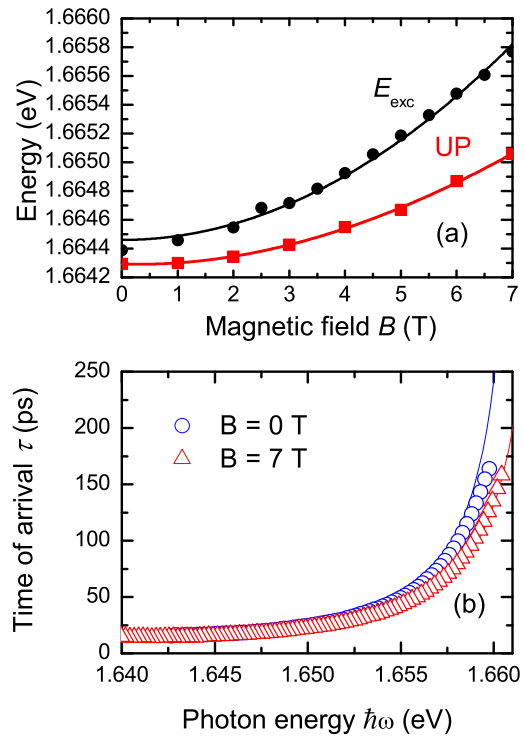}
\caption{(a) Magnetic field dependence of the energies of the exciton resonance $\hbar\omega_0$ and the UP, evaluated from reflectivity and PL spectra, respectively. Solid lines are fits with parabolas. (b) Time of arrival $\tau$ measured using a time-of-flight technique and plotted as function of photon energy at $B=0$ and $B=7$~T. Solid lines are calculated using Eqs.~\eqref{eq:n_group} and \eqref{eq:kappa} with the average energy of the exciton resonance $\hbar\omega_0=1.6644$~eV at $B=0$ and $\hbar\omega_0=$1.6658~eV  at $B=7$~T. The inhomogeneous and homogeneous linewidths are $\hbar \Gamma_{\rm inh}= 1$~meV and $\hbar \Gamma=4~\mu$eV, respectively. \label{fig:diamagnetic}}
\end{figure}

The magnetic field induces a diamagnetic shift of exciton spectra towards higher energies.
Figure \ref{fig:diamagnetic}(a) shows the magnetic field dependence of the exciton resonance and the UP-peak energies evaluated from reflection and PL spectra, respectively. The energies of both features grow quadratically with magnetic field.
For the exciton we deduce a diamagnetic shift of $28~\mu$eV/T$^2$, while for the UP-peak the value is somewhat smaller and equals to $15.8~\mu$eV/T$^2$.  The latter is close to the one of previously measured diamagnetic shifts of $14.7~\mu$eV/T$^2$ for excitons bound to acceptors in CdTe crystals.\cite{Molva83}

The diamagnetic shift influences the time-of-flight measurements.
The time of arrival $\tau$ increases with decreasing detuning energy $\delta$.
An increase in the energy of the exciton resonance increases the detuning and, consequently reduces $\tau$.
This shift is in accordance with the diamagnetic shift of $E_{exc}=\hbar\omega_0$ in the reflectivity spectra, see Fig.~\ref{fig:diamagnetic}(b).
We cannot resolve the Zeeman splitting of the exciton-polariton sublevels due to significant inhomogeneous broadening.
However, as we show below, this information can be evaluated through the polarimetric time-of-flight measurements.

\section{\label{sec:TOF} Polarimetric time-of-flight technique}

\begin{figure*}
\includegraphics[width=17cm]{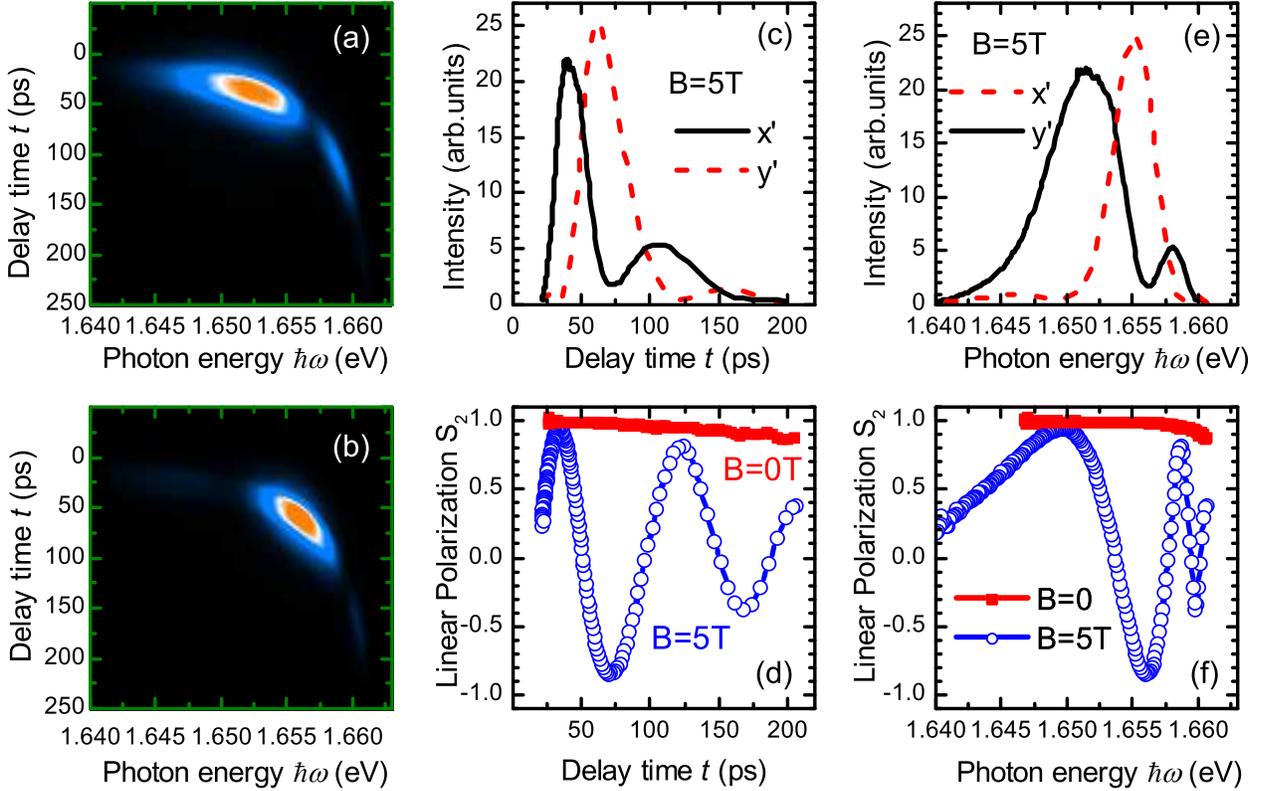}
\caption{(a) and (b) Contour plots of transmitted pulse intensities measured as function of photon energy $\hbar\omega$ and time delay $t$ at $B=5$~T. Panels (a) and (b) correspond to the data measured in linear $x^\prime$ and $y^\prime$ polarizations, respectively. (c) gives temporal dependencies of the intensities $I_{x^\prime}$ (solid line) and $I_{y^\prime}$ (dashed line) and (d) shows the temporal evolutions of linear polarization degree $S_2$ measured at $B=5$~T [open circles] and $B=0$ (solid squares). The spectral dependences of $I_{x^\prime}$ and $I_{y^\prime}$  are displayed in (e) and of $S_2$ in (f). The incoming pulse is linearly polarized along $x^\prime$ ($S_2=1$).
\label{fig:pola-TOF}}
\end{figure*}

Figure \ref{fig:pola-TOF} summarizes the temporal and spectral variations of the transmitted optical pulse intensity and its linear polarization in the axes frame $x'$, $y'$, $S_2$, acquired using the polarimetric time-of-flight technique. The use of 10~meV spectrally broad pulses centered at 1.656~eV allows us to monitor the exciton-polariton dispersion in a single measurement.
The results of two such measurements, where the transmitted intensity taken for linear polarizations along the $x^\prime$ and $y^\prime$ axes at $B=5$~T, are shown by the contour plots in Figs.~\ref{fig:pola-TOF}(a) and \ref{fig:pola-TOF}(b).
The initial polarization of the optical pulse is along $x^\prime$ axis ($S_{2}=1$). The time $t=0$ corresponds to the arrival time of the optical pulse without the sample in the optical path.

Strong temporal distortion of the optical pulse takes place after transmission through the 655~$\mu$m thick sample, see Fig.~\ref{fig:pola-TOF}(c). The full width at half maximum is about 100~ps and the pulse tail approaches delays as high as $200-300$~ps. This is a result of the strong exciton-polariton dispersion, which manifests as a significant increase of the group index $n_g$, when $\hbar\omega$ approaches the energy of the exciton resonance $\hbar\omega_0$. The group index $n_g$ and the resulting arrival time $\tau$ are independent of the initial polarization within the accuracy of experiment.
As demonstrated in Fig.~\ref{fig:diamagnetic}(b), they are well described using the dielectric function $\varkappa(\omega)$ in Eq.~\eqref{eq:kappa} with $\hbar\omega_0=1.6644$~eV, $\hbar \Gamma_{\rm inh}=1$~meV and $\hbar \Gamma=4~\mu$eV as obtained from reflection and transmission spectra discussed in the previous section.
In addition, for $B>0$ the diamagnetic shift of the exciton resonance should be taken into account in accordance with the data in Fig.~\ref{fig:diamagnetic}.

Application of a magnetic field leads to pronounced oscillations in the temporal and spectral dependences of the linear polarization $S_2$ as shown in Figs.~\ref{fig:pola-TOF}(d) and \ref{fig:pola-TOF}(f) respectively. To avoid confusion we stress here that already at the photon energies of about $1$~eV a very strong Faraday rotation (amounting to about $45^\circ$ at $B=5$~T) is observed. Hence, even at the smallest delays the polarization plane of transmitted light is rotated with respect to the polarization of the incident beam. This effect is discussed in detail below, in Sec.~\ref{sec:circ-lin}.
This is in contrast to the zero magnetic field data, also shown in Fig.~\ref{fig:pola-TOF}.
In zero magnetic field the polarization state of the incoming pulse is fully conserved.
One can see that $S_2$ is almost constant against the time delay and in the spectral domain.
First, this means that our system is isotropic at $B=0$.
Second, nearly no depolarization takes place, so that the exciton-polariton propagation is coherent.

The frequency of the oscillations induced by a magnetic field of $B=5$~T  increases with an increase of $\hbar\omega$ in the spectral domain, see Fig.~\ref{fig:pola-TOF}(f).
In the time domain, shown in Fig.~\ref{fig:pola-TOF}(d), it is almost constant.
This is in accord with our expectations for a magnetic field induced anisotropy due to splitting of exciton Zeeman sublevels.
Indeed, as follows from Eq.~\eqref{eq:phi-group} the phase difference between the two eigenwaves $\phi \approx \Omega_{\rm eff}\tau$ increases linearly with time of arrival and the resulting polarization degree oscillates in the time domain with frequency $\Omega_{\rm eff}$.
This allows us to determine the effective Zeeman splitting, even when it is hidden by inhomogeneous broadening in the spectral domain.
Here the polarimetric time-of-flight experiment shows similarities to time-resolved PL or pump-probe Faraday rotation techniques with which the Larmor precession of the spin can be resolved directly in time.\cite{Akimov09}
We observe more than two oscillations of linear polarization in a time window of 200~ps at $B=5$~T.
Such strong retardation results from the large delays of light in the sample.
This result is quite interesting from the application point of view, because it allows to achieve optical modulation at high frequencies of about 10~GHz.

\begin{figure}
\includegraphics[width=\linewidth]{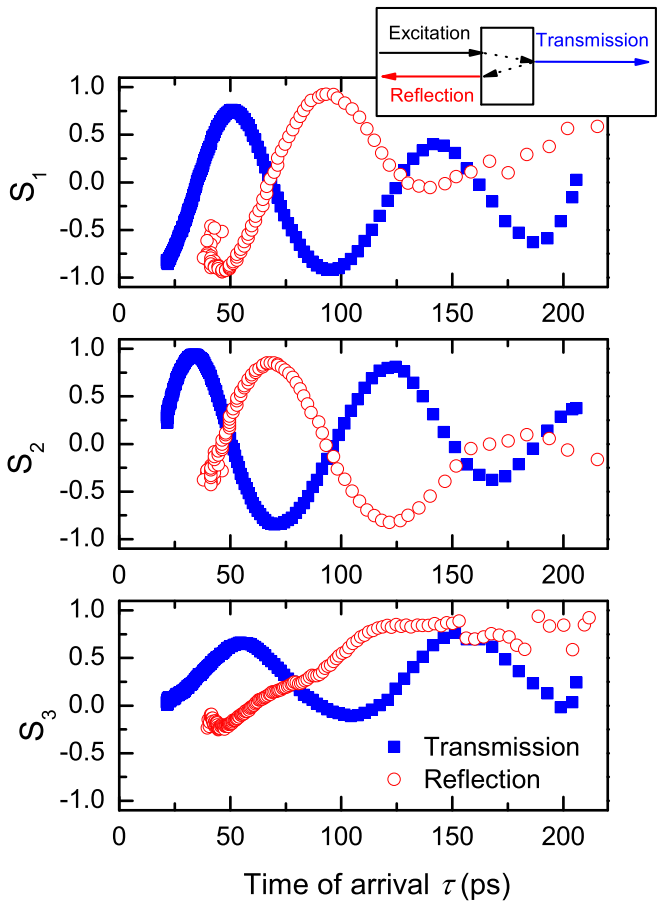}
\caption{Time dependence of Stokes parameters measured in transmission (blue squares) and reflection (red circles) geometry. In reflection geometry the light is reflected at the back side and travels through the crystal twice. The data are taken at $B=5$~T and the incoming light polarization $S_2=1$.
\label{fig:Reflection-TOF}}
\end{figure}

The temporal [Fig.~\ref{fig:pola-TOF}(d)] and spectral [Fig.~\ref{fig:pola-TOF}(f)] dependencies of Stokes parameters are mutually linked through the exciton-polariton dispersion. The data at larger delays in time domain directly correspond to the spectral components of the optical pulse with smaller detuning $\delta$ and vice versa. However, there are certain differences which make temporal measurements more attractive. First, as already mentioned above, the oscillation frequency in time directly corresponds to the splitting of exciton Zeeman sublevels. Second, the spectral dependence is influenced by inhomogenous broadening of the exciton resonance, while in temporal domain for large detuning ($\delta \gg \Delta,\hbar\Gamma$) the inhomogeneity of the group index cancels out.

The time evolutions of all three Stokes parameters are shown in Fig.~\ref{fig:Reflection-TOF}.
The optical pulse is linearly polarized along $x^\prime$ ($S_2=1$) before entering the sample.
The Stokes parameters are measured in two configurations: transmission and reflection.
In the second case we analyze the signal, which corresponds to the reflection of light from the back side of the sample.
This means that the exciton-polaritons perform a round trip in the crystal.
Subsequently, the outgoing reflection is detected with a streak camera as shown in the inset of Fig.~\ref{fig:Reflection-TOF}.
Note that in a time-of-flight measurement the identification of beams due to reflection of light from the front and back surface is straightforward, since each of the reflections has a characteristic time of arrival.

There are two main results, which follow from Fig.~\ref{fig:Reflection-TOF}.
First, in the transmission geometry all three Stokes parameters oscillate synchronously.
It is obvious, that the appearance of circular polarization $S_3$ and, even more important, its oscillations cannot be explained in terms of the Faraday effect only. The latter provides only continuous conversion between the linear polarizations described by the Stokes parameters $S_1$ and $S_2$, i.e., a rotation of polarization plane.
Second, the data in transmission and reflection geometries differ strongly. The transmitted light shows strong oscillations in all three polarizations while the reflected light shows oscillations only in the two linear polarizations ($S_1$ and $S_2$).
In circular polarization we only see a drift towards positive circular polarization.

Both these results evidence that in addition to Faraday rotation there is significant contribution due to non-reciprocal birefringence.
First, this non-reciprocal birefringence is responsible for conversion between the $S_2$ and $S_3$ polarizations. Second, it also changes sign when the direction of propagation is inverted, see Eq.~\eqref{disp}.  For the light that makes a round trip in the sample the linear birefringence, therefore, tends to compensate itself on the way back. This is consistent with the strong differences in time evolution of the circular polarization $S_3$ in different geometries and the disappearance of its oscillatory behavior in reflection. In the next Section we show that the ratio between Faraday rotation and non-reciprocal birefringence exhibits a strong spectral dependence resulting in nutation of exciton-polariton polarization.

\section{\label{sec:circ-lin} Circular versus linear birefringence}

We have shown above that both Faraday rotation (circular birefringence) as well as non-reciprocal birefringence (linear birefringence) play important roles in the polarization evolution of exciton-polaritons.
The Faraday rotation is responsible for transformation between the $S_1$ and $S_2$ Stokes parameters, while the non-reciprocal birefringence governs the transfer between $S_2$ and $S_3$.
Surprisingly, although non-reciprocal birefringence involves both wavevector and magnetic field effects, see Ref.~\onlinecite{Burstein71} and our Eq.~\eqref{disp}, it is strong enough in our sample to be observed in the Faraday geometry.

\begin{figure}
\includegraphics[width=\linewidth]{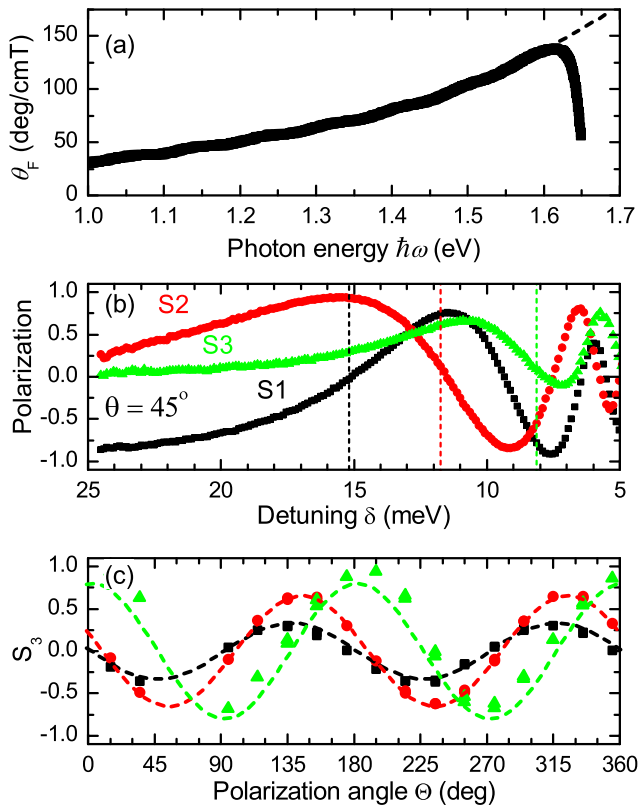}
\caption{(a) Faraday rotation over a wide spectral range below the exciton resonance obtained using time-integrated measurements with a halogen lamp. Dashed line is the linear extrapolation of the experimental data for the region of 1.5$-$1.6~eV used to estimate the contribution $\gamma_1^\prime$ from higher energy resonances to the Faraday rotation. (b) Spectral dependence of the Stokes parameters in vicinity of the exciton resonance measured in transmission geometry. The data are taken at $B=5$~T and the initial polarization $S_2=1$. (c) Degree of circular polarization $S_3$ as function of the polarization angle $\Theta$ at $B=5$~T and three different photon energies with $\delta = 15.2$ (black squares), 11.7 (red circles), and 8.1~meV (green triangles) marked in (a) by dashed vertical lines. Solid lines are fits with Eq.~\ref{eq:S3}.
\label{fig:spectral}}
\end{figure}

In order to separate the different contributions we first analyze the spectral dependence of the Stokes parameters after transmission of initially linearly polarized light ($S_2=1$). This data is shown in Fig.~\ref{fig:spectral}(b) and corresponds to Fig.~\ref{fig:Reflection-TOF} in the time domain. From these data it follows that ellipticity and, in particular, circular polarization $S_3$ appear only when the photon energy approaches the exciton resonances.
Thus for large detunings $\delta>25$~meV the non-reciprocal birefringence is negligible, while the Faraday rotation is still large.
Figure~\ref{fig:spectral}(a) shows that Faraday rotation takes place in a much wider spectral range even at photon energies below 1~eV. Moreover, its spectral dependence is not monotonic. It grows up to 1.61~eV and then abruptly drops down when approaching the exciton resonance. Such kind of dependence is result of two contributions. The first one is spectrally broad and originates from energetically higher lying interband optical transitions (with 3$-$4~eV energy) in the $X$ and $L$ points of the Brillouin zone.\cite{Cade1985}
These transitions exhibit very strong oscillator strength due to high density of states.\cite{Furdyna04} The second contribution is spectrally narrow and originates from the exciton resonance as discussed in Section \ref{subsec:micro}. These two contributions have different signs, which leads to the non-monotonic dependence of Faraday rotation in Fig.~\ref{fig:spectral}(a).
As a result, the Faraday rotation cancels at a certain photon energy. Close to that energy the relative strengths of the magneto-optical effects $q$, introduced in Eq. (\ref{neff}), can change drastically.

As it follows from Fig.~\ref{fig:spectral}(b), which shows the transmitted polarization as a function of detuning $\delta$, at the detuning energy  $\delta=15.2$~meV Faraday rotation is absent. Indeed, the exciting light was $S_2=1$ polarized and since no $S_1$ polarization was generated at this energy there is no Faraday effect.
Further insight into the strengths of Faraday rotation and non-reciprocal birefringence can be obtained from the rotational anisotropy. Figure~\ref{fig:spectral}(c) shows the dependence of the circular polarization degree $S_3$ as function of the angle $\Theta$ between the polarization plane of the initially linearly polarized light and the $x$-axis.
The data are shown for three different detunings.
At $\delta=15.2$~meV there is no Faraday rotation and we observe pure linear birefringence in the basis related to the $x$ and $y$ axes.
The conversion of polarization from $S_2$ to $S_3$ is most efficient at $\Theta=\pi/4+m\pi/2$ and it is zero at $\Theta=m\pi/2$ ($m=0,\pm1,\pm2...$).
If both non-reciprocal birefringence and Faraday rotation are present the degree of circular polarization $S_3$ in the transparency region can be written as
\begin{equation}
\label{eq:S3}
S_3=\frac{q}{\sqrt{1+q^2}}\sin({2\Theta})\sin{(\phi)}+\frac{2q}{1+q^2}\cos{(2\Theta)}\sin^2{(\phi/2)}.
\end{equation}

Without Faraday rotation only the first term in Eq.~\eqref{eq:S3} is present ($q\rightarrow\infty$) and $S_3=\sin{(2\Theta)}\sin{(\phi)}$, which is the case for $\delta=15.2$~meV.
At this detuning energy the conversion has a maximum at $\Theta=\pi/4$ and oscillates with the phase difference of the eigenwaves $\phi$.
The combination of Faraday effect and non-reciprocal birefringence leads to a shift of the angle for maximum conversion due to the second term of Eq.~(\ref{eq:S3}). This is in full accord with the experimental data presented in Fig.~\ref{fig:spectral}(c), where the $S_3$ dependence experiences a phase shift for smaller detuning energies.

As expected, the strength of the non-reciprocal birefringence increases with decreasing $\delta$.
At $\delta>10$~meV the dependence $S_3(\Theta)$ is symmetric with respect to zero.
However, for smaller $\delta$ we observe an asymmetry along the direction of positive $S_3$ [see triangles in Fig.~\ref{fig:spectral}(c)].
This is  due to circular dichroism which is not accounted for in Eq.~\eqref{eq:S3}.
Nevertheless, in case of pure non-reciprocal birefringence ($\delta = 15.2$~meV) the dichroism can be neglected.
One can extract a phase difference of $\phi=0.32$ using Eq.~\eqref{eq:S3} from the data in Fig.~\ref{fig:spectral}(c).
Using Eq.~\eqref{eq:phi-group} to calculate the effective splitting $\Omega_{\rm eff}$, which is $\hbar \Omega_{\rm eff} = 2 V k_z B$ due to the absence of Faraday rotation, we evaluate $V = \phi c / [2 (n_g -n_0) k_z L B_z]= 5 \times 10^{-12}$~eVcm$\textup{T}^{-1}$,  using Eq.~\eqref{eq:phi-group}.
We compare this quantity with the theoretical estimate according to Eq.~\eqref{tildeV:KB} for the parameters of CdTe\cite{Landolt-41b} which gives $|V|= 1.3 \times 10^{-12}$~eV~cm~T$^{-1}$. The factor of $\sim 4$ difference may result from the difference of material parameters and the simplified form of the 8-band model used here.

A non-reciprocal birefringence for pure CdTe for detunings between $63$ and $440$~meV was measured in Ref.~\onlinecite{Krichevtsov99-CdTe}.
There it is characterized, in accordance with Refs.~\onlinecite{Gogolin84-GaAs,Krichevtsov98-CdMnTe}, by the parameter $A$.
It is related to $\gamma_2$ introduced in Eq.~\eqref{subeq:gamma} by $A=\gamma_2/2$.
The detunings in our study are smaller than $25$~meV, but Eq.~\eqref{subeq:gamma} allows us to extrapolate $\gamma_2$ for larger detunings.
We find that the extrapolated values for Cd$_{0.88}$Zn$_{0.12}$Te are by a factor of two smaller the measured values for CdTe.
This discrepancy can be due to the different experimental techniques or the presence of Zn in our sample.

\begin{figure}
\includegraphics[width=7cm]{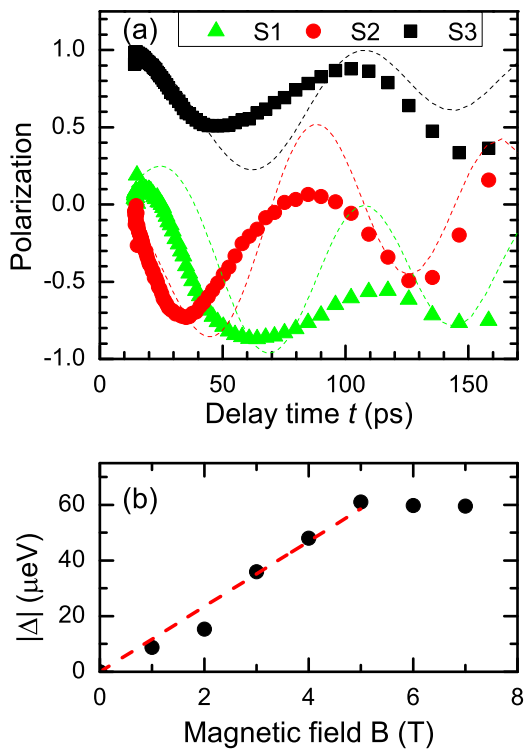}
\caption{(a) Time evolution of Stokes parameters at $B=7$~T for initially circularly polarized light ($S_3=1$). Dashed lines are fits with Eq.~\eqref{eq:phi} using the Zeeman splitting of the exciton spin sublevels $\Delta$ as the only fitting parameter. (b) Magnetic field dependence of $\Delta$ evaluated from the fits in (a). From a linear fit (dashed line) of the data in the range up to $B=5$~T we evaluate the exciton g-factor $|g_{\rm exc}|=0.2$.
\label{fig:B-dep}}
\end{figure}

The spectral dependence of the relative strength of Faraday rotation and non-reciprocal birefringence $q$ results in nutation of the light polarization in the  time domain. As a result, the Stokes parameters do not follow a pure time-periodic behavior, as would be the case for constant $q$. In contrast, we observe a continuous shift of the Stokes parameters, which manifests strongest when the incoming light is circularly polarized ($S_3=1$). The data are shown in Fig.~\ref{fig:B-dep}(a).

In order to accomplish a quantitative description of the magnetic-field-induced optical anisotropy in (Cd,Zn)Te we need to describe the Faraday rotation, which is sensitive to the Zeeman splitting of the exciton and the higher energy transitions.
The latter varies very slowly with photon energy in the region of interest ($\delta<25$~meV).
Therefore, it is sufficient to account for the Faraday rotation due to higher energy optical transitions by a constant $\vartheta_F^\prime$.
Its value of $\vartheta_F^\prime=158$ deg/(cm T), which corresponds to $\gamma_1^\prime=2.2\times  10^{-4}$~T$^{-1}$, is estimated by linear extrapolation [see the dashed line in Fig.~\ref{fig:spectral}(a)].

The total Faraday rotation includes two contributions, resulting both from high energy transitions and the exciton transition, the latter is given by Eq.~\eqref{subeq:gamma} convoluted with a Gaussian distribution in analogy to Eq.~\eqref{eq:kappa}.
The same convolution is performed for the non-linear birefringence contribution $\gamma_2$ [see Eq.~\eqref{subeq:gamma}] to take the inhomogeneous broadening into account. Finally the phase difference $\phi=\frac{\omega}{c}L(n_+-n_-)$ can be calculated using Eq.~\eqref{neff}. All quantities except the Zeeman splitting of the exciton resonance $\Delta$ are known.
The consistent modeling of all three Stokes parameters in the time domain using $\Delta$ as the only fitting parameter gives a good agreement with experiment [dashed curves in Fig.~\ref{fig:B-dep}(a)]. Using this procedure for different $B$ we determine the magnetic field dependence of the Zeeman splitting, which is shown in Fig.~\ref{fig:B-dep}(b). As expected the Zeeman splitting grows linearly with increasing magnetic field in the range $B<5$~T. From the linear dependence we evaluate the exciton $g$-factor to be $|g_{\rm exc}|=0.2$.
This value can be compared with the exciton $g$-factor in bulk CdTe estimated taking into account the complex valence band effects.\cite{Altarelli:1973ys,Cho75,Venghaus:1979vn} Calculation gives $|g_{\rm exc}|\approx 0.1$ if the electron-hole exchange interaction is disregarded\cite{Venghaus:1979vn} or $|g_{\rm exc}|\approx 0.35$ with allowance for electron-hole exchange. In these estimations we used an electron $g$-factor $g_e=-1.64$ and included the renormalization of the magnetic Luttinger parameter $\varkappa$ from $0.35$ to $0.25$ owing to magnetic field induced mixing of exciton states. We stress that the strong sensitivity of the $g$-factors to the band structure parameters and Zn contents makes a more precise comparison unwarranted.
The deviation of Zeeman splitting $\Delta$ from linear dependence at magnetic fields $B>5$~T is mainly due to complex structure of the valence band.

\section{\label{sec:Conclu} Conclusions}

We have studied the polarization dynamics of exciton-polaritons in sub-mm thick (Cd,Zn)Te bulk crystals subject to a longitudinal magnetic field (Faraday geometry). The strong decrease of the group velocity close to the exciton resonance leads to a sub-ns delay of light depending on the photon energy. This results in a significant enhancement of magneto-optical effects, which manifests in oscillations of the polarization state in the time domain when a spectrally broad pulse is transmitted through the crystal. The characteristic frequency depends on magnetic field and reaches 10~GHz at $B\sim5$~T.
Our data provide a novel approach for the determination of the exciton Zeeman splitting between the exciton sublevels using time-of-flight technique. It is especially useful if the energy splitting is small and hidden by inhomogeneous broadening, so that it cannot be resolved by linear spectroscopic techniques.

In addition to the frequently observed Faraday rotation (circular birefringence), we observe a significant contribution of non-reciprocal birefringence (linear birefringence). This observation occurs due to the non trivial spectral dependence of the Faraday rotation, which changes its sign at photon energies close to the exciton resonance.
As a result, each spectral component of the optical pulse experiences a different anisotropy due to the steep energy dependence of the relative strength of Faraday rotation and non-reciprocal birefringence, leading to polarization nutation of the exciton-polaritons.
A microscopic, model which accounts for Faraday rotation and magneto-spatial dispersion, provides a good quantitative description of the experimental data.
We evaluate the exciton $g$-factor $|g_{\rm exc}|=0.2$ and the magneto-spatial dispersion term $V k_z B_z$ with $V = 5 \times 10^{-12}$~eV cm $\textup{T}^{-1}$. Nutation of the exciton-polariton polarization is a quite interesting phenomenon, because it  can be applied to manipulation of the polarization state of the transmitted light. In this case the relative strength of linear and circular birefringence can be adjusted by active control of the energy of the exciton resonance.

\section{\label{sec:Ack} Acknowledgements}

The authors are grateful to H.J.~Weber for useful discussions and X-ray diffraction measurements. The work was supported by the Deutsche Forschungsgemeinschaft through grants AK40/4-1 and BA 1549/19-1  and by the Bundesministerium f\"{u}r Bildung und Forschung through the grant 05K12PE1. MMG is grateful RFBR and RF President Grant NSh-5442.2012.2.


\begin{references}
\bibitem{note1} {The expressions for $K$-linear terms of Refs.~\onlinecite{Gogolin84-GaAs} and \onlinecite{pikusmarushaktitkov} differ by the factor $4/\sqrt{3}$, here we use the notations of Ref.~\onlinecite{pikusmarushaktitkov}.}
\bibitem{note2} A full treatment would require at least a 14-band model and will be reported elsewhere.
\end{references}
\end{document}